\documentclass{ws-rv9x6}
\usepackage{subfigure}     
\usepackage{ws-rv-van} 
\newcommand{\vecr}{{\bf r}}

\newcommand{\vecR}{{\bf R}}

\newcommand{\Vee}{{\sf V}}
\newcommand{\brho}{{\mbox{\boldmath $\rho$}}}
\newcommand{\half}{\frac{1}{2}}

\begin{document}

\chapter{Reaction mechanisms of pair transfer}\label{ptm}

\author{Ian J. Thompson}

\address{Lawrence Livermore National Laboratory,
\\ PO Box 808, L-414,  Livermore, CA 94551, USA}

\begin{abstract}
The mechanisms of nuclear transfer reactions are described for the transfer of two nucleons from one nucleus to another. Two-nucleon
overlap functions are defined in various coordinate systems, and their transformation coefficients given between coordinate systems. 
Post and prior couplings are defined for sequential transfer mechanisms, and it is demonstrated that the combination of `prior-post' couplings 
avoids non-orthogonality terms, but does not avoid couplings that do not have good zero-range approximations. The simultaneous and sequential mechanisms are demonstrated for the  $^{124}$Sn(p,t)$^{122}$Sn reaction at 25 MeV using shell-model overlap functions. 
The interference between the various simultaneous and sequential amplitudes is shown.
\end{abstract}

\body

\section{Introduction}
Much of the evidence for nucleonic pairing in nuclei comes from energy expectation values,
but important further information comes from the {\em transfer} of pairs of nucleons to or from another nucleus of known structure.
In this regard, a more fundamental understating of nuclear reactions has been, and will continue to be (especially in the FRIB era), crucial to the nuclear physics community. This chapter focuses on the theory, calculation and model results for the reactions mechanisms of pair transfer.

Here we consider the reaction mechanisms for pair transfer between two nuclei, namely reactions that we can describe 
as $A(B{+}2,B)A{+}2$. Here, the two nucleons may be two neutrons, two protons, or a proton and a neutron, and are transferred from core $B$ to core $A$. The nucleons may transfer either in one {\em simultaneous} step, or one after the other {\em sequentially}. If a distinguishable proton and a neutron are transferred, then both proton-then-neutron and neutron-then-proton routes need to be considered. Furthermore, these sequential and simultaneous routes contribute  amplitudes that all add together {\em coherently}. This feature enables us to probe the nature of coherent two-nucleon superpositions in nuclei. Conversely, these superpositions, coupling orders and phase conventions have all to be defined consistently in a good calculation. 

Subsequent sections will therefore consider the definition of two-nucleon overlap functions, their coordinate transformations, 
the definition of transfer matrix elements along with zero-range approximations and non-orthogonality corrections. Finally, some results are shown to illustrate the coherence effects in the reaction mechanisms of pair transfers.

In the last 50 years, a significant number of papers have been presented in which absolute differential cross sections have been calculated, and compared with experimental results 
\cite{Yoshida1962685,Glendenning65,Bjerregaard1966145,Ascuitto69, Ascuitto197117, Bayman19711,Broglia197360,Charlton76,Yasue:1977p4589,Hashimoto78,Takemasa1979269,Bayman82,Maglione198559,Kurokawa:1987p4586,Igarashi:1991p4591,Tanihata08,Potel:2009p4441,Potel10,Potel11}. 
Traditionally, for example in \cite{Yasue:1977p4589}, the theory predictions have fallen well below the experimental data. This ratio has been called the `unhappiness factor' \cite{Mermaz-PhysRevC.20.2130,Vaagen1979}, and has sometimes been more than 100.
Most previous calculations modeled the transfer of a dineutron as a single cluster. And only from 
Charlton \cite{Charlton76} were sequential transfer contributions considered. 
We find that modern calculations, such as as those of Potel {\em et al}\cite{Potel11}, are in considerably better agreement with experiment.

\section{Bound states and vertex functions}
\label{wf+vertex}
The general theory of nucleon pair bound states defines the overlap function 
$\phi^{J}_{I} (\vecr , {\brho} )  = \langle \Phi_A(I) | \Phi_{A+2}(J)\rangle$ 
in terms of the Jacobi coordinates $\vecr$ between the two nucleons, 
and $\brho$ between their center of mass (cm) and the core $A$. 
The core spin is $I$ and the spin of the $A{+}2$ composite state is $J$.
When intrinsic spins $s_1, s_2$ are also included in a particular coupling order such as 
$ \left | \{L , (\ell , (s_1 s_2) S ) j \}    J_{12} , I ;~J \right\rangle $,
we have the partial wave expansion
\begin{eqnarray} \nonumber
  \phi^{J}_{I} (\vecr , \brho ) =\!\!\!
    \sum_{ L  \ell S j  J_{12} I} \!\!\!\!\!\!
     && 
       \phi_{I \mu_I} (\xi_c)  
       \phi_{s_1}^{\sigma_1} \phi_{s_2}^{\sigma_2} ~
        Y_{L}^{\Lambda} ( \hat\vecr ) 
        Y_\ell^\mu (  \hat {\brho} ) ~
        {1 \over r \rho} u_{12} (r, \rho) \langle J_{12} M_{12}  I \mu_I | J M\rangle
\\  \label{tntwf}
    && 
       \langle L \Lambda j m_{12} | J_{12} M_{12}\rangle
       \langle \ell \mu S \Sigma         | j m_{12}\rangle
       \langle s_1 \sigma_1 s_2 \sigma_2  | S \Sigma\rangle \ .
\end{eqnarray}
The radial wave function $u_{12} (r, \rho) $
can be given either as a cluster product of single-particle wave functions
$ u_{12} (r, \rho)  =\Phi_L (r)\phi_\ell (\rho), $
input directly as a two-dimensional distribution e.g. from a Faddeev bound-state
calculation, or calculated from the
correlated sum of products of single-particle states with independent coordinates. 
These two-nucleon wave functions will in general be the eigenstates of a three-body 
bound state Schr\"odinger equation
\begin{eqnarray}
  [ T_\vecr + T_\brho + V_{1A} + V_{2A} + V_{12} - \varepsilon ] \phi^{J}_{I} (\vecr , \brho )  = 0\ ,
\end{eqnarray}
where the $V_{iA}$ are the potentials between nucleon $i$ and the core, and $V_{12}$ is the
pairing interaction between the two nucleons.

Such two-particle states that come from shell-model calculations \cite{Cohen:1970p4595}  or from Sturmian-basis
calculations \cite{Bang:1985p483} are usually described by means of the
$ | \vecr_1 , \vecr_2\rangle $ coordinates. 
This describes a pair state as
\begin{eqnarray} \label{indepwf}
  \varphi_{12} ( \vecr_1 , \vecr_2 ) &=&
   \sum _ i c_i ~
   \left | ( \ell_1 (i), s_1 )j_1 (i),
            ( \ell_2 (i), s_2 )j_2 (i);~ J_{12}  \right\rangle
\end{eqnarray}
The coefficients $c_i$ for correlated basis states $i$ and the single-particle wave functions $\varphi_{\ell s j}(\vecr)$
contain all the physics information about the bound state needed to do a transfer calculation. Shell model codes \cite{nushell}
can produce the coefficients $c_i$ needed here in terms of previously calculated eigenstates of the $A$ and the $A+2$ systems. 
These states are then transformed into the centre-of-mass coordinates
$ | \vecr ,  {\brho} \rangle $ of Eq. (\ref{tntwf})
using $ \vecr_i = x_i \vecr + y_i {\brho} $.
For equal mass particles, $x_1 = x_2 = 1 $, and
$y_1 = - y_2 =\half $.

The {\em vertex functions} of these bounds states are defined to be these bound state wave functions $\phi^{J}_{I} (\vecr , \brho ) $
multiplied by the potentials which have zero effects after the transfer step is performed
and all exit channel nuclei have completely separated. These potentials are therefore the sum $V$
of the binding potentials $V^{\rm sp}_i = V_{\ell s j}(\vecr_i)$,
namely $V = V^{\rm sp}_{1A} + V^{\rm sp}_{2B}$.
(These are the individual potentials that should appear  in the bound-state equation 
$[ T_\vecr + V_{\ell s j}(\vecr) - \varepsilon] \varphi_{\ell s j}(\vecr) = 0$.)
The vertex function does {\em not} include the nucleon-nucleon 
pair interaction $V_{12}(\vecr_1{-}\vecr_2)$, since this potential produces binding effects in both the initial and final bound states. We denote 
by $V \phi^{J}_{I} (\vecr , \brho ) $ the vertex function after transformation into Jacobi coordinates by the same method 
used to transform the wave function itself.

\section{Post and prior coupling forms of transfer matrix elements}

We now consider the Hamiltonian $\cal H$ for the whole system of $A{+}B{+}2$ nucleons and
described by system wave function $\Psi$ for the complete transfer reaction  $A(B{+}2,B)A{+}2$.
Let the various $A{+}2$ and $B{+}2$ bound states be denoted by $\Phi_i$ for indices $i$. Then we may expand $\Psi$ in terms of the $\Phi_i$ with some
coefficients $\psi_i ( \vecR_i ) $ depending on the two-body separation vectors $\vecR_i$. This gives the channel expansion 
 $\Psi  = \sum_i \psi_i ( \vecR_i )  \Phi_i$.

The model Schr\"odinger's equation $ [ {\cal H} - E ]\Psi = 0$ when projected
separately onto the different basis states $\Phi_j , $ yields the set of equations
\begin{eqnarray} \label{CRC1}
\left [ E_j - H_j \right ]\psi_j ( \vecR_j ) +
   \sum _ {i \neq j}
         \left\langle\Phi_j | {\cal H}-E |\Phi_i \right\rangle
         \psi_i ( \vecR_i ) =0,
\end{eqnarray}
which couple together the unknown wave functions $\psi_i (\vecR_i).$
The channel Hamiltonians are defined by the diagonal $H_j - E_j = \langle\Phi_j | {\cal H}-E |\Phi_j \rangle$
such that the $E_j$ are the asymptotic kinetic energies in channel $j$.

The off-diagonal matrix element $\langle\Phi_j | {\cal H} - E |\Phi_i\rangle $
has two different forms, depending on whether we expand
\begin{eqnarray}
{\cal H} - E & =& H_j - E_j + V_j \mbox{ (the `post' form)} \nonumber \\
             & =& H_i - E_i + V_i \mbox{ (the `prior' form)}.\nonumber
\end{eqnarray}
The name (post or prior) is determined by whether it is the initial or final channel whose Hamiltonian is used.
The above Eq. (\ref{CRC1}), as written, has $i$ as the initial channel and $j$ as the final channel
for the indicated coupling.
Thus
\begin{eqnarray} \label{popr}
 \left\langle\Phi_j | {\cal H} - E |\Phi_i \right\rangle & = &
   V_{ji}^{\rm post} + [H_j - E_j ] K_{ji} {\rm ~~~(post)} \\
  {\rm or } & =& V_{ji}^{\rm prior} + K_{ji} [H_i - E_i ]
                      {\rm ~~~(prior),}  \nonumber
\end{eqnarray}
where
\begin{eqnarray}
  V_{ji}^{\rm post}  \equiv\langle\Phi_j | V_j |\Phi_i \rangle , ~~~
   V_{ji}^{\rm prior}   \equiv\langle\Phi_j | V_i |\Phi_i\rangle , ~~~
   K_{ji}              \equiv\langle\Phi_j | \Phi_i\rangle . \label{overlaps}
\end{eqnarray}
The overlap function $K_{ji} =\langle\Phi_j |\Phi_i\rangle $ in
Eqs.\ (\ref{popr},\ref{overlaps}) arises from the  non-orthogonality between the
basis states $\Phi_i $ and $\Phi_j $ if these are in
different mass partitions. The $K_{ji}$ are non-local operators that go to zero asymptotically ($R_i$ or $R_j \to \infty$).
(Within the {\em same} partition, the $\Phi_i$ would be inelastic states, and would form an orthogonal set.)

The {\em first-order DWBA} matrix element use entrance $\psi_i$ and exit $\psi_j$ channel wave functions
satisfying $[H_i - E_i] \psi_i = 0$ and $[H_j - E_j]\psi_j = 0$ respectively.
Its matrix element is
\begin{eqnarray}
T^{(1)}_{ji} &=& \langle \psi^{(-)}_j \Phi_j | {\cal H} - E |  \Phi_i \psi^{(+)}_i \rangle \ .
\end{eqnarray}
The prior form of this is 
\begin{eqnarray}
T^{(prior)}_{ji} &=& \langle \psi^{(-)}_j \Phi_j | H_i - E_i + V_i  |  \Phi_i \psi^{(+)}_i \rangle \nonumber \\
                &=& \langle \psi^{(-)}_j \Phi_j |  V_i  |  \Phi_i \psi^{(+)}_i \rangle + \langle \psi^{(-)}_j \Phi_j |    \Phi_i [H_i - E_i] \psi^{(+)}_i \rangle \nonumber \\
                &=& \langle \psi^{(-)}_j \Phi_j |  V_i  |  \Phi_i \psi^{(+)}_i \rangle + 0\nonumber \\
                &=& \langle \psi^{(-)}_j  |   V_{ji}^{\rm prior}   |  \psi^{(+)}_i \rangle\ .
\end{eqnarray}
Similarly, the equivalent post form is 
\begin{eqnarray}
T^{(post)}_{ji} &=& \langle \psi^{(-)}_j \Phi_j | H_j - E_j + V_j  |  \Phi_i \psi^{(+)}_i \rangle \nonumber \\
                &=& \langle \psi^{(-)}_j \Phi_j |  V_i  |  \Phi_i \psi^{(+)}_i \rangle + \langle \psi^{(-)}_j  [H_j - E_j] \Phi_j |    \Phi_i\psi^{(+)}_i \rangle \nonumber \\
                &=& \langle \psi^{(-)}_j \Phi_j |  V_j  |  \Phi_i \psi^{(+)}_i \rangle  + 0\nonumber \\
                &=& \langle \psi^{(-)}_j  |   V_{ji}^{\rm post}   |  \psi^{(+)}_i \rangle\ .
\end{eqnarray}
Thus the non-orthogonality term disappears in first-order DWBA.
Post and prior first-order DWBA matrix elements can be made to exactly agree numerically, if sufficient care is taken to ensure convergence of the non-local form factors.

Let a {\em second-order DWBA} matrix element use entrance channel $i$, exit channel $k$, and some intermediate channel $j$, as $i \to j \to k$.
The propagation in  the intermediate channel may be described in terms of the Green's function $G_j$, 
or equivalently within an iterated coupled-channels set.  
The two-step DWBA matrix element is 
\begin{eqnarray}
T^{(2)}_{ki} &=& 
      \langle \psi^{(-)}_k |~\langle\Phi_k | {\cal H} {-} E  |  \Phi_j  \rangle G_j 
      \langle \Phi_j | {\cal H} {-} E  |  \Phi_i \rangle~ |\psi^{(+)}_i \rangle \ .
\end{eqnarray}
Now there are {\em four} matrix elements that may be calculated, 
according to the first and the second Hamiltonian form chosen: post-post, post-prior, prior-post, and prior-prior. The terms prior and post for each step are used to refer to the initial or final channels {\em of that step}, not the overall incoming or outgoing channels. In `prior-post', the prior refers to the first step, and the post refers to the second step.

The  post-post form of this, for example, is 
\begin{eqnarray}
T^{(post,post)}_{ki} = 
      \langle \psi^{(-)}_k |~\langle\Phi_k | H_k {-} E_k {+} V_k  |  \Phi_j \rangle G_j 
       \langle \Phi_j | H_j {-} E_j {+} V_j |  \Phi_i \rangle~| \psi^{(+)}_i \rangle.~~~~
\end{eqnarray}
Here the $[H_k-E_k]$ can operate on the final $\psi_k$ to give zero, but little can simplify the $[H_j-E_j]$ since 
$[H_j-E_j]G_j \ne 0$ always. Thus
\begin{eqnarray}
T^{(post,post)}_{ki} &=&
      \langle \psi^{(-)}_k | V_{kj}^{\rm post} G_j   V_{ki}^{\rm post} |\psi^{(+)}_i \rangle  + \nonumber \\
       	  &&    \langle \psi^{(-)}_k | V_{kj}^{\rm post} G_j  [ H_j - E_j ]  K_{ji} |~\psi^{(+)}_i \rangle .
\end{eqnarray}
This second term is called a `non-orthogonality term' since it involves the bound-state non-orthogonality overlaps $K_{ji}=\langle\Phi_j | \Phi_i\rangle$, which is significant when $R_i$ and $R_j$
are both within the range of the bound states.

Similar analyses for post-prior and prior-prior two-step DWBA expression also have non-orthogonality terms in the final form. 
The {\em prior-post form}, however, is 
\begin{eqnarray}
T^{(prior,post)}_{ki} = 
      \langle \psi^{(-)}_k |~\langle\Phi_k | H_k {-} E_k {+} V_k  |  \Phi_j \rangle G_j 
       \langle \Phi_j | H_i {-} E_i {+} V_i |  \Phi_i \rangle~|\psi^{(+)}_i \rangle.~~~~
\end{eqnarray}
Here  the $[H_i-E_i]$ can also operate on the initial $\psi_i$ to give zero, 
as well as $[H_k-E_k]$ on the $\psi_k$,
so we have the simplest form
\begin{eqnarray}
T^{(prior,post)}_{ki} &=&
      \langle \psi^{(-)}_k | V_{kj}^{\rm post} G_j   V_{ji}^{\rm prior} |\psi^{(+)}_i \rangle  \ .
\end{eqnarray}
The non-orthogonality terms can  thus be made to disappear in second-order DWBA if
the first and second steps use the prior and post interactions respectively.
If the non-orthogonality terms are included as necessary, the results should be the same whatever post or prior forms are used.

In third and higher-order transfer calculations, some non-orthogonality terms will always be present, but most pair transfer mechanisms can be well modeled as two-step processes.

\section{Two-nucleon transfer interaction}
\label{NN-transfers}
We now consider the specific transfer matrix element 
$V_{ji}^{\rm prior} = \langle\Phi_j | V_i |\Phi_i \rangle$.
Given an expression for this prior form, we may calculate the post interaction easily as $V_{ji}^{\rm post} = (V_{ij}^{\rm prior})^\dagger$.
Take $\Phi_j$ to refer to the bound states of nucleus $A{+}2$ outside core $A$, and $\Phi_i$ analogously for nucleus $B$.

The transfer interaction has therefore the non-local matrix element
\begin{eqnarray} \label{transfer-me}
  \Vee_{ji}(\vecR_j, \vecR_i) = \langle \phi^{J_A}_{I_A} (\vecr , \brho_A ) | 
      V^{\rm sp}_{1B}+  V^{\rm sp}_{2B}+ U_{AB}  - U_i| 
      \phi^{J_B}_{I_B} (\vecr, \brho_B ) \rangle .
\end{eqnarray}
As is usual in transfer operators, there are three kinds of potentials appearing here.
First there are the binding potentials $V^{\rm sp}_{1B}(\vecr_{1B}) +  V^{\rm sp}_{2B}(\vecr_{2B})$.
Since these binding potentials always appear while multiplied by their bound state wave functions, we need only store and use
the vertex functions defined in section \ref{wf+vertex}.
Secondly, there is the `core-core' potential $U_{AB}(\vecR_{AB}) $ between the core nuclei $A$ and $B$.
Finally is subtracted an optical potential. In this prior form we subtract the optical potential in the initial channel, $U_i(\vecR_i)$.
The difference $U_{AB}  - U_i$ of the two optical potentials is called the {\em remnant term}, and is sometimes taken to be small.

The integrals in Eq. (\ref{transfer-me}) include integrating over the two-nucleon separation $\vecr$
as well as over their cm distance $\brho_A$ from the core $A$.
 The $\vecr$ coordinate appears  in both the initial and final states, and so is not labeled by $A$ or $B$. This has the important consequence that neither the distance nor the angle of the $\vecr$ coordinate is changed in the transfer. Neither, therefore, is their relative angular momentum $\ell$, and, for similar reasons, nor  their spin couplings $S$ and total angular momentum $j$.
The two neutron transfer can hence be viewed as the transfer of a
`structured particle'
$\{  r, ( \ell , (s_1 s_2) S ) j  \}$, and then becomes similar to the more familiar single-particle transfers.
This means that when we also afterwards integrate over the coordinate $\brho_A $, we can use the standard procedures already
developed for one-particle transfer interactions.

\section{Coordinate transformations}
\label{transformations}
The transfer mechanism requires the pair wave function to be expressed in the form of Eq. (\ref{tntwf}), so independent-particle
forms of Eq. (\ref{indepwf}) have to be transformed in their coordinates as 
\begin{eqnarray}
  \varphi_{12} ( \vecr_1 , \vecr_2 ) =  
  \sum _ u {c_i} \sum_{ L \ell S j}
   \left | L , ( \ell , (s_1 s_2 )S )j ;
                         J_{12} T \right\rangle
 \phi^{J_{12} T,i}_{L(\ell S)j} (r, \rho)\ .
\end{eqnarray}
A particular basis state $i$ in the $(r,\rho)$ coordinates is
\begin{eqnarray}
\phi^{J_{12},i}_{L(\ell S)j} (r, \rho) &=&
  \left\langle L , ( \ell , (s_1 s_2 )S )j ;J_{12} \right |
  \left ( \ell_1 (i), s_1 )j_1 (i),
       ( \ell_2 (i), s_2 )j_2 (i);~ J_{12} \right\rangle \nonumber
\\
&& \times\langle
   \left [  Y_L (\hat \vecr) Y_\ell (\hat \rho) \right ] _ \lambda
    |
   \left [ \varphi_{\ell_1 s_1 j_1} (\vecr_1)\varphi_{\ell_2 s_2 j_2} (\vecr_2)
   \right ] _ {J_{12} T}  \rangle
\end{eqnarray}
where (suppressing the $i$ indices for clarity), and including an isospin $T$ to define the antisymmetrization,
\begin{eqnarray} \nonumber
&&\!\!\!\langle L , ( \ell , (s_1 s_2 )S )j ;J_{12} T |
       ( \ell_1 , s_1 )j_1 ,
       ( \ell_2 , s_2 )j_2 ;~ J_{12} T\rangle = \sum _ \lambda  \hat \lambda \hat S \hat {j_1} \hat {j_2}
~~~~\\
&&   \left ( \begin{array}{ccc}\ell_1 & \ell_2 & \lambda\\
                            s_1    & s_2    & S \\
                            j_1    & j_2    & J_{12} \end{array}  \right )
         {1 + (-1)^{\ell +S+T} \over
    \sqrt {2(1 + \delta_{\ell_1 , \ell_2} \delta_{ j_1 ,j_2}) }} ~
\hat j \hat \lambda W(L \ell J_{12} S; \lambda j) (-1)^{\ell+L-\lambda} .~~~
\end{eqnarray}
The radial overlap integral can be derived by means of harmonic-oscillator
expansions \cite{mosh60}, with the Bayman-Kallio expansion \cite{bay67}
or using the Moshinsky solid-harmonic expansion\cite{Moshinsky1959104}. This last method gives
\begin{eqnarray}
 &&K^\lambda_{\ell L: \ell_1 \ell_2} (r, \rho) =\langle
   \left [  Y_L (\hat\vecr) Y_\ell (\hat{\brho}) \right ] _ \lambda
    |
   \left [ \varphi_{\ell_1} (\vecr_1)\varphi_{\ell_2} (\vecr_2)
   \right ] ^ \lambda \rangle  \\
 &=  &\sum_{n_1 n_2} ~
      \left ( \begin{array}{c} 2 \ell_1 {+}1\\2n_1\end{array} \right ) ^\half
      \left ( \begin{array}{c} 2 \ell_2 {+}1\\2n_2\end{array} \right ) ^\half
      (x_1 r)^{\ell_1 - n_1}
      (y_1 \rho)^{n_1}
      (x_2 r)^{n_2}
      (y_2 \rho)^{\ell_2 - n_2} \nonumber
\\
&&\times \sum _ Q
      {\bf q}_{\ell_1 \ell_2}^Q (r, \rho) ~
     (2Q{+}1)~ \hat {\ell_1} \hat {\ell_2} \widehat {\ell_1 {-} n_1} \widehat {\ell_2 {-} n_2} ~
             \hat L \hat \ell  \nonumber
\\
&&\times  \sum_{\Lambda_1 \Lambda_2}
       \left ( \begin{array}{ccc}\ell_1 {-} n_1&n_2&\Lambda_1\\0&0&0\end{array} \right )
       \left ( \begin{array}{ccc}\ell_w {-} n_2&n_1&\Lambda_2\\0&0&0\end{array} \right )
       \left ( \begin{array}{ccc}\Lambda_1&L   &Q\\0&0&0\end{array} \right )
       \left ( \begin{array}{ccc}\Lambda_2&\ell&Q\\0&0&0\end{array} \right )  \nonumber
\\
&&\times (-1)^{\ell_1 + \ell_2 +L+ \Lambda_2} (2 \Lambda_1 + 1) (2 \Lambda_2 + 1)
        W(\Lambda_1 L \Lambda_2 \ell; Q \lambda) \nonumber
       \\ 
&&     \times   \left ( \begin{array}{ccc} \ell_1{-}n_1& n_2& \Lambda_1\\
                                   n_1       & \ell_2{-}n_2& \Lambda_2\\
                                   \ell_1    & \ell_2 & \lambda \end{array} \right )  ,
\end{eqnarray}
where $ \left ( \begin{array}{c}a \\ b \end{array}\right ) $ is a binomial coefficient.
%
The kernel function
${\bf q}_{\ell_1 \ell_2}^Q (r, \rho) $ is the Legendre expansion of the product of
the two radial wave functions in terms of $u$,
the cosine of the angle between $\vecr$ and $ {\brho} $:
\begin{eqnarray}
 {\bf q}^Q_ {\ell_1 , \ell_2} (r, \rho)
 =\half
\int_ {-1}^{+1}
         {\varphi_{\ell_1s_1j_1} (r_1) \over r_1}^{\ell_1 +1}
       {\varphi_{\ell_2s_2j_2} (r_2)\over r_2}^{\ell_2 +1} ~
         P_Q (u) du
\end{eqnarray}

\section{Zero-range and other approximations}
The coupling potentials $\Vee_{ji}(\vecR_j, \vecR_i) $ of Eq.\ (\ref{transfer-me}) are {\em non-local}, in the sense that in general the initial and final radii, $\vecR_j$ and $\vecR_i$, will be different. They will not only have different magnitudes, but also different directions.
In the early days of transfer modeling, the calculations only became practical if a {\em zero-range} approximation could be found, in which the coupling was restricted to $\vecR_j = \alpha \vecR_i$  for some constant $\alpha$ (which need not be unity).

When the projectile is a light ion such as $^3$H, $^3$He or $^4$He for nucleus $B+2$, then the binding potential sum $V^{\rm sp}_{1B}+  V^{\rm sp}_{2B}$ will have short range. We may therefore consider approximating the vertex function 
\begin{eqnarray}
  [V^{\rm sp}_{1B}+  V^{\rm sp}_{2B} ]  \phi^{J_B}_{I_B} (\vecr, \brho_B ) \sim D_{0} \delta(\brho_B) \phi^B_{nn}(\vecr) 
\end{eqnarray}
for some nucleon-nucleon wave function $\phi_{nn}(\vecr)$ that we are free to choose. This a zero-range approximation. Note that it is only $\brho_B$ which needs to have zero range, not $\vecr$. The constant $D_{0}$ is called the {\em zero-range constant}.

If, furthermore, we can neglect the remnant term $U_{AB}  - U_i$, then the transfer coupling of Eq.\ (\ref{transfer-me}) 
can be simplified as
\begin{eqnarray} \label{transfer-zr}
  \Vee_{ji}(\vecR_j, \vecR_i) &=& \langle \phi^{J_A}_{I_A} (\vecr , \brho_A ) | 
      V^{\rm sp}_{1B}+  V^{\rm sp}_{2B} | \phi^{J_B}_{I_B} (\vecr, \brho_B ) \rangle  \nonumber \\
    &=&   \langle \phi^{J_A}_{I_A} (\vecr , \brho_A ) |   D_0 \delta(\brho_B) \phi^B_{nn}(\vecr)  \rangle \nonumber \\
    &=&  D_0 ~  \langle \phi^B_{nn}(\vecr) | \phi^{J_A}_{I_A} (\vecr , \brho_A ) \rangle ~~  \delta(\brho_B)  \nonumber \\
    &=&  D_0 ~  \langle \phi^B_{nn}(\vecr) | \phi^{J_A}_{I_A} (\vecr , \brho_A ) \rangle 
          ~~ \delta (\beta \left(\vecR_j -  \frac{A}{A{+}2}  \vecR_i \right)  ),
\end{eqnarray}
since 
\begin{eqnarray} 
\vecR_j -  \frac{A}{A{+}2}  \vecR_i =  \brho_B / \beta \mbox{~~~for~~~} \beta = \frac{2(A{+}B{+}2)}{(A{+}2)(B{+}2)}.
\end{eqnarray} 
That is, we arrive at a  `form factor' $\langle \phi^B_{nn}(\vecr) | \phi^{J_A}_{I_A} (\vecr , -\vecR_j  ) \rangle$ that is local in
$\vecR_j = \frac{A}{A+2}  \vecR_i = - \brho_A$ because of the delta function $\delta(\brho_B)$. To find the form factor, we need to determine the average nucleon-nucleon relative wave function $\phi^B_{nn}(\vecr)$ in the light ion, and project the heavy-nucleus two-body wave function $\phi^{J_A}_{I_A} (\vecr , \brho_A )$ onto this relative motion. This gives a function only of the distance $\rho_A = R_j$ and the angles. The kinematics for this zero-range approximation are identical to those for the one-body transfer of a mass-2 cluster from core $B$ to core $A$. A local-energy approximation may be used to improve the treatment of the finite range of the vertex function, just as for one-body transfers.

This is a further example the conclusion stated at the end of section \ref{NN-transfers}, namely that transfer reactions only probe in the unknown nucleus those components of $nn$ relative motion that already exist in the known nucleus. Since the known light nuclei
$^3$H, $^3$He and $^4$He have predominantly $s$-wave relative motion between the two transferred nucleons, our transfer reactions will only probe pairing states of $s$-wave relative motion in the target. The {\em magnitude} of the transfer cross section will be proportional to the form factor overlap $\langle \phi^B_{nn}(\vecr) | \phi^{J_A}_{I_A} (\vecr , \brho_A ) \rangle$.

Zero-range approximations can be also used for some of the sequential steps involving these light nuclei, but not for all of them if we are using `prior-post' couplings to avoid non-orthogonality corrections. 
For stripping reactions such as (t,p), the first prior (t,d) step has no good
zero-range approximation, and for pickup reactions such as (p,t), the second post (d,t) step must be treated in full finite range for the same reason.




%

\begin{table}[tb]
\tbl{Two-neutron overlap function for $\langle ^{122}$Sn$| ^{124}$Sn$\rangle$}
{\begin{tabular} {lr}
\hline
$1g_{7/2}^2$ \hspace*{15mm}& 0.62944 \\ 
$2d_{5/2}^2$ & 0.59927 \\ 
$2d_{3/2}^2$ & 0.71913 \\ 
$3s_{1/2}^2$ & 0.51892 \\ 
$1h_{11/2}^2$ & --1.24399 \\ 
\hline
\end{tabular}}
\label{sn124-22}
\end{table}

\begin{figure}[tb]
\begin{center}
\psfig{file=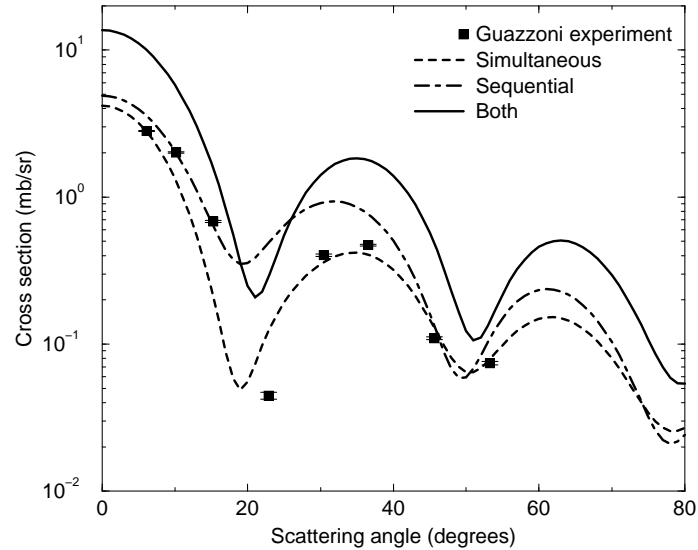,width=0.8\textwidth} 
\end{center}
\caption{Simultaneous (short dash), sequential (dot-dash) and simultaneous+sequential (solid line) cross sections for the reaction $^{124}$Sn(p,t)$^{122}$Sn at 25 MeV,
in comparison with the experimental data of Guazzoni {\em et al.} \cite{Guazzoni:2011p4491}.}
\label{sn124pt1}
\end{figure}

\begin{figure}[tb]
\begin{center}
\psfig{file=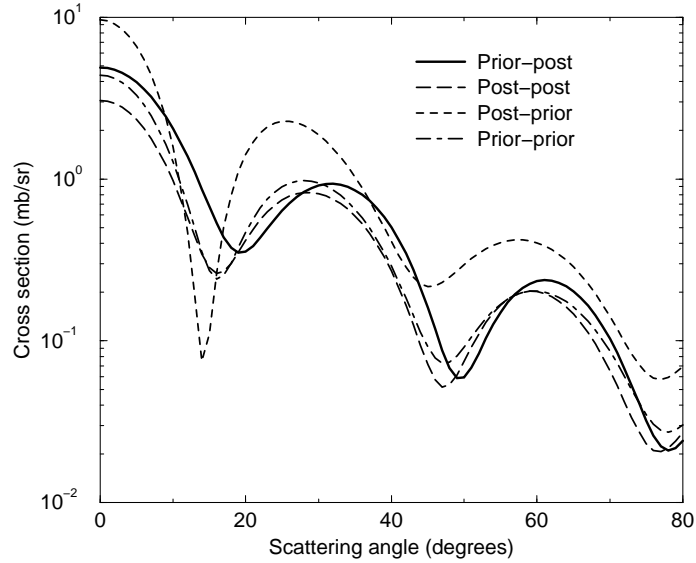,width=0.8\textwidth} 
\end{center}
\caption{Sequential cross sections for all possible combinations of post and prior for the two steps.}
\label{ppall}
\end{figure}

\begin{figure}[tb]
\begin{center}
\psfig{file=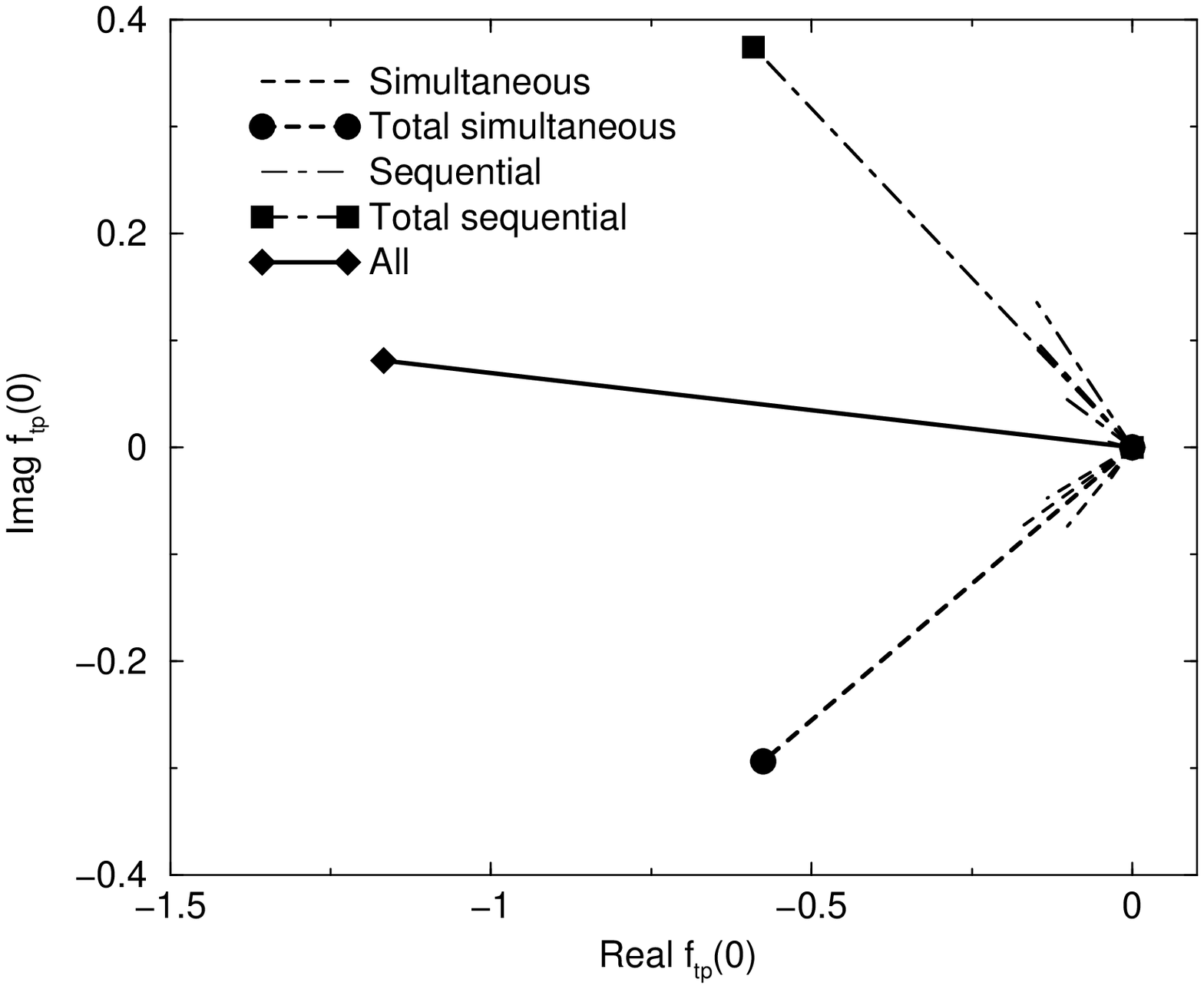,width=0.8\textwidth} 
\end{center}
\caption{Simultaneous (short dash), sequential (dot-dash) and simultaneous+sequential (solid line) amplitudes at zero degrees for the reaction $^{124}$Sn(p,t)$^{122}$Sn at 25 MeV. The short lines show the individual contributions from the wave function
components of Table \ref{sn124-22}, and the longer lines with symbols are their coherent sums.}
\label{sn124ptamps}
\end{figure}

\section{Results}
In this short paper we will focus on the 
reaction mechanisms for the pair transfer $^{124}$Sn(p,t)$^{122}$Sn at 25 MeV, using the overlap function shown in Table \ref{sn124-22} we find by overlapping the shell-model wave functions for the ground states of $^{122}$Sn and $^{124}$Sn. 
The structure results for $^{124}$Sn and $^{122}$Sn were obtained
in the model space of 
$  (0g_{7/2}$, $1d_{5/2}$, $1d_{3/2}$, $2s_{1/2}$, $0h_{11/2})  $
for neutrons with the code NuShell \cite{nushell}.
The model-space
two-body matrix elements are those used in Refs. \cite{mom1,mom2}.
They were obtained starting with a G matrix
derived from the CD-Bonn \cite{bonn} nucleon-nucleon interaction.
The harmonic oscillator basis was employed for the
radial wave functions with an oscillator energy $\hbar\omega$ =
7.87 MeV. The effective interaction for the above shell-model
space is obtained from the $  Q  $-box method and includes all
non-folded diagrams through third-order in the interaction G, 
to sum up the folded diagrams to infinite order \cite{mbpt1,mbpt2}.
The single-particle energies were adjusted to reproduce
the observed states in $^{131}$Sn.

The inputs to the
reaction code are the two-nucleon spectroscopic amplitudes (TNA) of Table \ref{sn124-22}.
A center of mass correction \cite{cm}
equal to $  [A/(A{-}2)]^{2n+\ell }  $
for the TNA has been applied, where $  A=124  $.
Our sign convention is that the radial wave functions
are positive at the origin.
The sequential process was calculated
by a single intermediate state for each of these orbits
connected by a product of one-nucleon spectroscopic
amplitudes that are equal to the center-of-mass
corrected TNA multiplied by $  \sqrt{2}  $ that
takes into account the normalization of the two-particle
amplitude. Future calculations should also take into
account the TNA obtained from the mixing of neutron pairs
for orbitals outside of the model space.

We use the triton potential of Li \cite{Li2007103}, the deuteron potential of Daehnick \cite{Daehnick-PhysRevC.21.2253}, and the proton potential of Chapel Hill 89 \cite{Varner199157}. All the two-neutron wave functions are constructed within the half-separation-energy prescription.
For a triton wave function we use the pure $s^2$ configuration found by the product of eigenstates at the half-separation energy (4.24 MeV) in a Woods-Saxon potential with $V=77.83$ MeV, $R=0.95$ fm, and $a=0.65$ fm (the results are not sensitive to these values). 
The Sn wave functions shown in Table \ref{sn124-22} are found at the half-separation energy (7.219 MeV) in a WS potential with $r=1.17$ fm, and $a=0.75$ fm that has
the fixed spin-orbit component $V_{so}=6.2$ MeV, $r=1.01$ fm, and $a=0.75$ fm.

The complete cross section prediction is shown in Fig.\ \ref{sn124pt1}, compared with the experimental data of Guazzoni {\em et al.} \cite{Guazzoni:2011p4491}. Now we see that, with the shell-model overlaps and proper finite-range and sequential contributions, the  unhappiness factors are much closer to unity. A better agreement between theory and experiment has already been published\cite{Potel11}, but  in
the present calculations there are still questions about the angular oscillations which are in not so good agreement with experiment. Note that Guazzoni {\em et al.} \cite{Guazzoni:2011p4491} took the better agreement of the {\em simultaneous} transfer curve (dashed line) to indicate small effects for sequential transfers, but this is not correct since we do know that sequential transfers occur, and can calculate them with good accuracy in this model (dot-dashed line). 

To see the importance of the non-orthogonality terms, and hence of choosing `prior-post' couplings if non-orthogonality terms are to be avoided, Fig.\ \ref{ppall} plots the different sequential cross sections for all possible combinations of post and prior for the two steps. The prior-post solid curve is the dot-dashed curve in Fig.\ \ref{sn124pt1}. The other curves are all different from this one, and cannot be simply added to the simultaneous amplitude to get the correct result. This also implies that no complete calculation with only zero-range couplings is possible.

Finally, it is instructive to look at the interference effects between the various simultaneous and sequential contributions. To display these coherence effects, I choose to plot the scattering amplitude at zero degrees for the non-spin-flip amplitude $m_p = m_t = 1/2$ (the only non-zero amplitude at this angle). Fig.\ \ref{sn124ptamps} plots all the simultaneous and sequential contributions from the different 
components listed in Table \ref{sn124-22}, along with their coherent sums. We see that all the contributions to the simultaneous transfer are constructively coherent, as are all the contributions to the total sequential amplitude. This constructive coherence follows from the signs of the amplitudes in Table \ref{sn124-22}, and reflects the significant pairing enhancement in $^{124}$Sn.
The total sequential and simultaneous amplitudes are not uniformly coherent with each other, however. 
A uniform $90^\circ$ angle between the simultaneous and sequential amplitudes in Figure \ref{sn124ptamps} would indicate an incoherent summation of the two cross sections, but that is not exactly true either.  This  reflects the importance of the deuteron channel with its own specific optical potential. In general, varying deuteron optical potentials and differing intermediate $Q$-values require that both simultaneous and sequential terms be explicitly calculated.

\section*{Acknowledgements}
This work was supported by the TORUS topical collaboration, and performed under the auspices of the U.S. Department of Energy by Lawrence Livermore National Laboratory under Contract DE-AC52-07NA27344. I thank Alex Brown (MSU) for many discussions, and for providing shell-model overlap functions.


\bibliographystyle{ws-rv-van}
\bibliography{two-nt,two-nt2,bab}

\end{document}